\def\et{$E_T$}                          
\def\met{\mbox{${\hbox{$E$\kern-0.6em\lower-.1ex\hbox{/}}}_T$}} 
\def\D0{D\O}                            
\def\ppbar{$p\overline{p} $}            

\def\etaj2lt{$|\eta_2| <$}
\def\gtetaj2lt{$< |\eta_2| <$}
\documentstyle[preprint,eqsecnum,psfig,aps]{revtex}
\begin{document}

\tightenlines
\title{
Color Coherent Radiation in Multijet Events
from \ppbar\ Collisions at $\sqrt{s} = 1.8$ TeV} 
%
%
\author{                                                                      
B.~Abbott,$^{29}$                                                             
M.~Abolins,$^{26}$                                                            
B.S.~Acharya,$^{44}$                                                          
I.~Adam,$^{12}$                                                               
D.L.~Adams,$^{38}$                                                            
M.~Adams,$^{17}$                                                              
S.~Ahn,$^{14}$                                                                
H.~Aihara,$^{22}$                                                             
G.A.~Alves,$^{10}$                                                            
E.~Amidi,$^{30}$
% ...
\\                                                                            
\vskip 0.50cm                                                                 
\centerline{(D\O\ Collaboration)}                                             
\vskip 0.50cm                                                                 
}                                                                             
\address{                                                                     
\centerline{$^{1}$Universidad de los Andes, Bogot\'{a}, Colombia}             
\centerline{$^{2}$University of Arizona, Tucson, Arizona 85721}               
\centerline{$^{3}$Boston University, Boston, Massachusetts 02215}             
\centerline{$^{4}$Brookhaven National Laboratory, Upton, New York 11973}      
\centerline{$^{5}$Brown University, Providence, Rhode Island 02912}           
\centerline{$^{6}$Universidad de Buenos Aires, Buenos Aires, Argentina}       
\centerline{$^{7}$University of California, Davis, California 95616}          
\centerline{$^{8}$University of California, Irvine, California 92697}         
\centerline{$^{9}$University of California, Riverside, California 92521}      
\centerline{$^{10}$LAFEX, Centro Brasileiro de Pesquisas F{\'\i}sicas,        
                  Rio de Janeiro, Brazil}                                     
\centerline{$^{11}$CINVESTAV, Mexico City, Mexico}                            
\centerline{$^{12}$Columbia University, New York, New York 10027}             
\centerline{$^{13}$Delhi University, Delhi, India 110007}                     
\centerline{$^{14}$Fermi National Accelerator Laboratory, Batavia,            
                   Illinois 60510}                                            
\centerline{$^{15}$Florida State University, Tallahassee, Florida 32306}      
\centerline{$^{16}$University of Hawaii, Honolulu, Hawaii 96822}              
\centerline{$^{17}$University of Illinois at Chicago, Chicago,                
                   Illinois 60607}                                            
\centerline{$^{18}$Indiana University, Bloomington, Indiana 47405}            
\centerline{$^{19}$Iowa State University, Ames, Iowa 50011}                   
\centerline{$^{20}$Korea University, Seoul, Korea}                            
\centerline{$^{21}$Kyungsung University, Pusan, Korea}                        
\centerline{$^{22}$Lawrence Berkeley National Laboratory and University of    
                   California, Berkeley, California 94720}                    
\centerline{$^{23}$Louisiana Tech University, Ruston, Louisiana 71272}        
\centerline{$^{24}$University of Maryland, College Park, Maryland 20742}      
\centerline{$^{25}$University of Michigan, Ann Arbor, Michigan 48109}         
\centerline{$^{26}$Michigan State University, East Lansing, Michigan 48824}   
\centerline{$^{27}$Moscow State University, Moscow, Russia}                   
\centerline{$^{28}$University of Nebraska, Lincoln, Nebraska 68588}           
\centerline{$^{29}$New York University, New York, New York 10003}             
\centerline{$^{30}$Northeastern University, Boston, Massachusetts 02115}      
\centerline{$^{31}$Northern Illinois University, DeKalb, Illinois 60115}      
\centerline{$^{32}$Northwestern University, Evanston, Illinois 60208}         
\centerline{$^{33}$University of Notre Dame, Notre Dame, Indiana 46556}       
\centerline{$^{34}$University of Oklahoma, Norman, Oklahoma 73019}            
\centerline{$^{35}$University of Panjab, Chandigarh 16-00-14, India}          
\centerline{$^{36}$Institute for High Energy Physics, 142-284 Protvino,       
                   Russia}                                                    
\centerline{$^{37}$Purdue University, West Lafayette, Indiana 47907}          
\centerline{$^{38}$Rice University, Houston, Texas 77005}                     
\centerline{$^{39}$Universidade do Estado do Rio de Janeiro, Brazil}          
\centerline{$^{40}$University of Rochester, Rochester, New York 14627}        
\centerline{$^{41}$CEA, DAPNIA/Service de Physique des Particules,            
                   CE-SACLAY, Gif-sur-Yvette, France}                         
\centerline{$^{42}$Seoul National University, Seoul, Korea}                   
\centerline{$^{43}$State University of New York, Stony Brook,                 
                   New York 11794}                                            
\centerline{$^{44}$Tata Institute of Fundamental Research,                    
                   Colaba, Mumbai 400005, India}                              
\centerline{$^{45}$University of Texas, Arlington, Texas 76019}               
\centerline{$^{46}$Texas A\&M University, College Station, Texas 77843}       
}                                                                             
\date{\today}
\maketitle
\begin{abstract}
    We report on a study of color coherence effects in $p\overline{p}$ 
collisions at a center of mass energy $\sqrt{s}=1.8$~TeV.  The data
were collected with the D\O\ detector during the 1992--1993 run of the Fermilab 
Tevatron Collider.
We observe the presence of initial--to--final state color interference 
with the spatial correlations between soft and hard jets in 
multijet events in the central and in forward pseudorapidity regions.  The
results are compared to Monte Carlo simulations with 
different color coherence implementations and to the predictions of
${\cal O}(\alpha_{s}^{3})$ QCD calculations.
\end{abstract}


\newpage

    Color coherence phenomena in the final state have been very well established
in $e^+e^-$ annihilations \cite{cc1,cc2,cc3,cc4,cc5,cc5a}, in what has been 
termed the ``string'' \cite{cc6} or ``drag'' \cite{cc7} effect.  Particle 
production in the region between the quark and antiquark jets in 
$e^+e^- \rightarrow q\overline{q}g$ events is suppressed.  In perturbative 
quantum chromodynamics (QCD) such effects arise from interference between the 
soft gluons radiated from the $q$, $\overline{q}$, and 
$g$.  While quantum mechanical interference effects are expected in QCD, it is
important to investigate whether such effects survive the nonperturbative
hadronization process, for a variety of reactions over a broad kinematic
range, as predicted by local parton--hadron duality\cite{cc7}.

    The study of coherence effects in hadron--hadron collisions is considerably
more subtle than that in $e^+e^-$ annihilations due to the presence of colored 
constituents in both the initial and final states.  The structure of multijet
events in hard processes is influenced by the underlying color configurations at
short distances.  During a hard interaction, color is transferred from one
parton to another and the color--connected partons act as color antennae.  
Examples of color flow diagrams are shown in
Fig.~1 for $q\overline{q}$ and $qg$ scattering.  In
Fig.~1a ($q\overline{q}$) the color system in which interference
occurs is entirely between the initial and final states, whereas in
Fig.~1b ($qg$) interference also occurs both in the initial and in
final states due to their explicit color connection.
Gluon radiation from the incoming and outgoing partons forms jets of hadrons 
around the direction of these
colored emitters.  The soft gluon radiation pattern accompanying any hard 
partonic system can be represented, to leading order in $1/N_c$ where 
$N_c$ is the number of colors, 
as a sum of contributions corresponding to the color--connected 
partons.  Within the perturbative calculations, this is a
direct consequence of interferences between the radiation of various 
color emitters, resulting in the QCD coherence 
effects \cite{cc7,cc8,cc9}.  
    
    Color coherence, which results in a suppression of soft gluon radiation in
the partonic cascade in certain regions of phase space, can be approximated by 
{\em Angular Ordering} (AO).  For the case of outgoing
partons, AO requires that the emission angles of soft gluons decrease
monotonically as the partonic
cascade evolves away from the hard process. The radiation is confined to a 
cone centered on the direction of one parton, and is
bounded by the direction of its color--connected partner.  Outside this region
the interference of different emission diagrams becomes destructive and the
azimuthally integrated amplitude vanishes to leading order.
Conversely the emission angles increase for the incoming partons as the
process develops from the initial hadrons to the hard subprocess.  Monte
Carlo simulations including coherence effects probabilistically by means of AO 
are available for both initial and final state evolutions.  Conventional 
perturbative QCD calculations  taken to sufficiently high order should in 
principle incorporate these effects to any accuracy.  Use of the latter 
approach is limited, however, due to the current lack of such calculations 
beyond order $\alpha_{s}^{3}$.

    Evidence for color coherence effects between initial and final states in 
\ppbar\ interactions has been previously published \cite{cc10} for the central 
pseudorapidity region.  In this paper we report on a new study of 
initial--to--final state color coherence phenomena in \ppbar\ interactions
which also extends the measurements into the untested forward region.  We 
compare our central and forward measurements to the predictions of 
${\cal O}(\alpha_{s}^{3})$ QCD matrix element calculations as well as to 
several parton--shower Monte Carlo (MC) simulations.  This is the first direct 
comparison of such measurements to analytic ${\cal O}(\alpha_{s}^{3})$ 
QCD predictions.

   The \D0\ detector is described in detail elsewhere \cite{cc11}.  This
analysis uses the uranium liquid--argon sampling calorimeter to measure jet
final states.  The \D0\ calorimeter has fine segmentation in both azimuthal
angle $\phi$, and in pseudorapidity $\eta=- \ln[\tan(\theta/2)]$, where 
$\theta$ is the polar angle of the jet with respect to the proton beam. 
It has hermetic coverage for $|\eta|<4$ with fractional transverse energy
$E_T$ resolution of $\sim 80\%/\sqrt{E_T\textup{(GeV)}}$ for jets.  

    The data sample for this analysis \cite{cc12}, representing an integrated
luminosity of 8~$pb^{-1}$, was collected during the 1992--1993 Tevatron
Collider run.  Events were selected using a three level trigger. The first 
level required a coincidence of two scintillator hodoscopes located 
on either side of the interaction region, to ensure an inelastic collision.  
The next stage required the transverse energy of at least three 
calorimeter towers ($0.2 \times 0.2$ in $\Delta\eta \times \Delta\phi)$ to 
exceed a 7~GeV threshold in the region $|\eta|<3.2$.  The surviving events 
were analyzed by an online processor farm where jets were reconstructed using 
a simplified version of the final jet finding algorithm and an event was 
recorded to tape if it had a jet with $E_T>85$~GeV.

    Jets were reconstructed offline using an iterative fixed--cone clustering 
algorithm with cone radius $R=\sqrt{(\Delta\eta)^2 + (\Delta\phi)^2}=0.5$.
Spurious jets from isolated noisy calorimeter cells and accelerator losses
were eliminated by loose cuts on the jet shape.  The $E_T$ of each jet was
corrected for offsets due to the underlying event, multiple \ppbar\ 
interactions, and noise; out--of--cone showering; and detector energy response 
as determined from the missing transverse energy balance of photon--jets 
events \cite{esca}.  Events were required to have a measured vertex within
50~cm of the detector center.

    The remaining events were required to have three or more reconstructed
jets.  Jets were ordered in \et\ and labeled $E_{T1}>E_{T2}>E_{T3}$.  We 
required $E_{T1}>115$~GeV to avoid any bias introduced by the trigger 
threshold, and the third jet to have $E_T>15$ GeV.
Color coherence effects are measured with the angular distribution
in $(\eta,\phi)$ space of the softer third jet around the second 
jet.  The polar variables $R$ and
$\beta=\tan^{-1}(\frac{{\mathrm sign}(\eta_2)\cdot\Delta\phi_{32}}%
{\Delta\eta_{32}})$
were used to locate the third jet in a search disk of 
$0.6<R<\frac{\pi}{2}$ around the second jet (Fig.~2).  
Here, $\eta_i$ and $\phi_i$ are the pseudorapidity and azimuthal angle of the 
$i^{th}$ jet, $\Delta\eta_{32}=\eta_3 - \eta_2$, and 
$\Delta\phi_{32}=\phi_3 - \phi_2$.
We define $\beta=0$ to point towards the beam nearest to the
second jet, and $\beta=\pi$ to point towards the farther beam.
We study the interference effects in regions $|\eta_2|<0.7$ and 
$0.7<|\eta_2|<1.5$.  The pseudorapidity of the leading jet was not constrained, 
but the first and second jets were required to be in opposite $\phi$ 
hemispheres, i.e. $\frac{\pi}{2}<\Delta\phi_{21}<\frac{3\pi}{2}$.
After all selection criteria, a sample of 9,048 (5,776) events remains in the
central (forward) region.

    The measured angular distributions (Fig.~3) are compared to 
the predictions of 
several MC simulations that differ in their implementation of color 
coherence.  We employ the parton--shower MC event generators 
{\small {ISAJET}} 7.13 \cite{cc13}, {\small {HERWIG}} 5.8 \cite{cc14}, and 
{\small {PYTHIA}} 5.7 \cite{cc16}, and
a partonic event generator, {\small {JETRAD}} 1.2 \cite{cc15} to calculate 
the ${\cal O}(\alpha_{s}^{3})$ QCD predictions.
The {\small {ISAJET}} generator uses an independent shower development model 
without any color coherence 
effects.  Both {\small {HERWIG}} and {\small {PYTHIA}} incorporate initial 
and final state color interference effects by means of the AO approximation of 
the parton cascades.  In {\small {PYTHIA}}, the AO constraint can be turned off.
{\small {HERWIG}} and {\small {PYTHIA}} each employ a phenomenological model 
to describe the hadronization process.  {\small {HERWIG}} uses the cluster 
hadronization model and
{\small {PYTHIA}} implements the Lund string fragmentation model, 
which are both supported by the observations of color coherence phenomena 
in $e^+e^-$ annihilations.  

    The shower--based MC simulations were performed at the particle
(hadron) level after the nonperturbative hadronization process, whereas the 
{\small {JETRAD}} predictions were at the parton level.  Detector
$\eta$ and energy resolution effects were included in all 
predictions.  The generated events were subsequently processed using the same 
criteria employed for analyzing the data. 

    The $\beta$ distributions, normalized to the total number of events, for 
the data and {\small {HERWIG}} and {\small {ISAJET}} predictions are shown in 
Fig.~3 for
both central and forward regions.  In each case the data peak
near $\beta=\pi$ and this enhancement is more dramatic for higher $\eta$. 
The shape of these 
distributions is sensitive not only to the process dynamics but also to
phase space effects resulting from our jet and event selection
criteria.  For both regions, {\small {HERWIG}} is in 
good agreement with the data, whereas {\small {ISAJET}} shows systematic
deviations from the observed distributions. 

    The ratios of the observed $\beta$ distributions relative to the MC 
predictions for both $\eta$ regions are shown in Fig.~4.  The 
data show a clear enhancement of events compared to {\small {ISAJET}} near the 
event plane (i.e., the plane defined by the directions of the second jet and 
the beam axis, $\beta=0,~\pi$) and a depletion in the transverse plane
($\beta=\frac{\pi}{2}$). 
This is consistent with the expectation from initial--to--final state color 
interference that the rate of soft jet emission around the event plane be 
enhanced with respect to the transverse plane.  The {\small {PYTHIA}}
predictions include string fragmentation.  Without AO the {\small {PYTHIA}} 
distributions are significantly different from the data, while with AO turned 
on there is much better agreement, although there are still some residual 
differences in the ``near beam'' region.  In addition, from the 
$\mathrm {\frac{\small {Data}}{\small {HERWIG}}}$ and 
$\mathrm {\frac{\small {Data}}{\small {JETRAD}}}$
$\beta$ distributions, we conclude that both the AO approximation and 
${\cal O}(\alpha_{s}^{3})$ QCD describe the coherence effects seen 
in data in both $\eta$ regions.  

    The main sources of uncertainty on the data $\beta$ distributions are 
summarized in Table~\ref{PL_TABLE1}.  Since we report event normalized 
distributions, any possible uncertainty on quantities that affect the overall 
rate of events is minimized.  
For both $\eta$ regions the statistical and systematic uncertainties are
comparable.  Sources of systematic uncertainty arise from the jet energy
calibration, a possible $\eta$ dependence of the jet reconstruction efficiency,
and small $\eta$ biases caused by the jet reconstruction algorithm.

    We have also examined three main sources of systematic uncertainty related 
to the MC predictions.   
Varying the jet energy resolution within its measured
uncertainties resulted in changes to the MC $\beta$ distributions 
of less than 2.6\% (2.8\%) in the central (forward) region.  The 
effect of using different
parton distribution functions ({\small {CTEQ2ML}}, {\small {CTEQ2MF}}, and 
{\small {CTEQ2MS}} \cite{cteq}) was 
examined with {\small {JETRAD}} and found to produce variations of less 
than 2.0\% in the central and forward regions.
The {\small {JETRAD}} $\beta$ distributions varied by less than 1.3\% (2.5\%)
in the central (forward) region when the renormalization and factorization 
scales varied from $E_{T1}/2$ to $2E_{T1}$.  

    Table~\ref{PL_TABLE2} shows the $\chi^2$ values of fits to the various 
${\mathrm \frac{\small {Data}}{\small {Monte Carlo}}}$ ratios with 
a constant for the central and forward regions as well as for the combined 
sample with $|\eta_2|<1.5$.
All uncorrelated systematic uncertainties (i.e., energy scale and 
$\eta$ bias corrections) in the data were added in quadrature with the 
statistical uncertainties and were included in the calculation of $\chi^2$.  
The uncorrelated systematic uncertainty due to jet energy resolution was also
added in quadrature with the statistical uncertainties of the MC
predictions.  From the $\chi^2$ values we conclude
that {\small {HERWIG}} and {\small {JETRAD}} agree best with our data for both
$\eta$ regions, whereas {\small {ISAJET}} clearly disagrees with the data.
In addition, from the data to {\small {PYTHIA}} comparisons we conclude that 
for the process under study, string fragmentation alone cannot accommodate the 
effects seen in the data.  The AO approximation is an element of parton-shower 
event generators that needs to be included if color coherence effects are to 
be modeled successfully.

    The data $\beta$ distributions were also compared to the predictions of 
{\small {HERWIG}} at the parton level (before the nonperturbative
hadronization stage).  
From the comparison of the ${\mathrm \frac{\small {Data}}{\small {HERWIG}}}$ 
ratios at the parton and particle level we conclude that the hadronization
effects as modeled by {\small {HERWIG}} are negligible and do not influence 
our results.

    We processed a limited {\small {HERWIG}} event
sample through a full {\small {GEANT}} based detector simulation \cite{geant}
to investigate any possible detector effects not accounted for in the
jet energy and $\eta$ resolutions.  From the comparison of the $\beta$
distributions at the particle and detector levels such residual detector 
effects were found to be negligible.

    We have presented a study of color coherent radiation in
multijet events in \ppbar\ collisions.  We have measured the spatial 
correlations between the second and third leading $E_T$ jets in
the central and in forward pseudorapidity regions.  Comparisons of the data 
distributions with various color coherence implementations demonstrate a 
strong presence of initial--to--final state interference.  Parton shower MC 
simulations that implement color interference by means of angular ordering 
reproduce the data angular distributions well, with {\small {HERWIG}} giving 
the best representation.  Striking differences are found between the data
and MC models that do not incorporate color coherence effects in the parton 
shower development.  Our results also indicate that coherence effects as 
predicted by a $2 \rightarrow 3$ partonic level calculation provide a good 
representation of our data, giving additional evidence supporting the validity
of the local parton--hadron duality hypothesis.

    We express our deep appreciation to V. Khoze and T. Sj\"{o}strand for 
numerous valuable discussions on this work.  We also thank T. Sj\"{o}strand 
for helping us with the color coherence implementations in {\small {PYTHIA}}.
We thank the staffs at Fermilab and collaborating institutions for their
contributions to this work, and acknowledge support from the 
Department of Energy and National Science Foundation (U.S.A.),  
Commissariat  \` a L'Energie Atomique (France), 
State Committee for Science and Technology and Ministry for Atomic 
Energy (Russia),
CNPq (Brazil),
Departments of Atomic Energy and Science and Education (India),
Colciencias (Colombia),
CONACyT (Mexico),
Ministry of Education and KOSEF (Korea),
and CONICET and UBACyT (Argentina).
%

\newpage
\begin{table}[!]
    \centering
    \caption[] {Major uncertainties in the data $\beta$ distributions.\\}
    \begin{tabular}{lcc}
    Source of uncertainty & \etaj2lt\ 0.7 & 0.7 \gtetaj2lt\ 1.5 \\ \hline
    Jet energy scale corrections &            2.2\%             & 3.4\% \\
    Jet $\eta$ bias correction &           1.2\%             & 1.7\% \\ 
    Jet reconstruction efficiency &                        & \\ 
     ~~Near beam                &       $+1.5$\%        & $+5.0$\% \\
     ~~Far beam                 &       $-0.5$\%        & $-2.0$\% \\
    Statistical error   &              3.2\%          & 4.3\% \\ 
    \end{tabular}
    \label{PL_TABLE1}
\end{table}
\bigskip
\bigskip

\begin{table}[!]
    \centering
    \caption[] {$\chi^2$ values for 8 degrees of freedom of fits to the various 
         $\mathrm {\frac{\small {Data}}{\small {Monte Carlo}}}$ ratios 
         of $\beta$ distributions with a constant  
         for the central, forward, and combined samples.
         Statistical and uncorrelated systematic uncertainties were included.\\}
    \begin{tabular}{lccc}                                  
    Event & \multicolumn{3}{c}{$\chi^2$} \\ \cline{2-4}
    Generator & \etaj2lt\ 0.7 & 0.7 \gtetaj2lt\ 1.5 & \etaj2lt\ 1.5 \\ \hline
    {\small {ISAJET}}    &    48.0   & 24.9 & 74.4\\ 
    {\small {HERWIG}}    &    5.1    & 7.7  & 7.2 \\ 
    {\small {JETRAD}}    &    7.9    & 7.9  & 7.2 \\ 
    {\small {PYTHIA}}    &           &      &     \\ 
     ~~String fragmentation, but no AO    &    77.2  & 62.9 & 126.0   \\ 
     ~~String fragmentation, and AO       &    12.7  & 11.3 & 19.3    \\
    \end{tabular}
    \vspace{0.3cm}
    \label{PL_TABLE2}
\end{table}

\newpage
{\centerline {FIGURE CAPTIONS}}
\vskip 0.5in
\noindent
{\bf Figure 1} Color flow diagrams for (a) $q\overline{q}$ and (b) $qg$
               scattering.

\vskip 0.5in
\noindent
{\bf Figure 2} Three jet event topology illustrating the search disk (shaded 
               area) for the softer third jet around the second jet.

\vskip 0.5in
\noindent
{\bf Figure 3} Comparisons of the data $\beta$ distributions to the 
               predictions of 
               {\small {ISAJET}} and {\small {HERWIG}} for (a), (b) central 
               region and (c), (d) forward region. The error bars 
               include statistical errors only.

\vskip 0.5in
\noindent
{\bf Figure 4} Ratio of $\beta$ distributions between data and the predictions 
               of:
               (a) {\small {ISAJET}}, (b) {\small {PYTHIA}} with AO off, 
               (c) {\small {PYTHIA}} with AO on, (d) {\small {HERWIG}}, 
               (e) {\small {JETRAD}} for the central region; and (f)-(j) for the
               forward region respectively. The error bars 
               include statistical and uncorrelated systematic uncertainties.

\begin{figure} 

\vskip 0.2in
{\centerline {\bf Figure 1}}
\vskip 2in

 \hspace{6.5cm}
 \psfig{figure=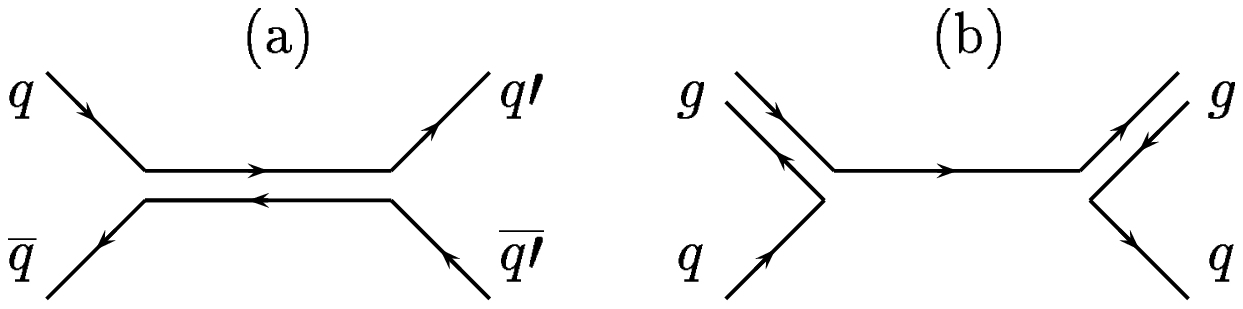,height=4.5cm,width=6.0cm}
 \vspace{-.5cm}
 \label{PL_FIG1}
 \vspace{0.3cm}
\end{figure}

\newpage
\vskip 0.5in
{\centerline {\bf Figure 2}}
\vskip 2in

\begin{figure} 
 \hspace{-1.5cm}
 \centerline{\psfig{figure=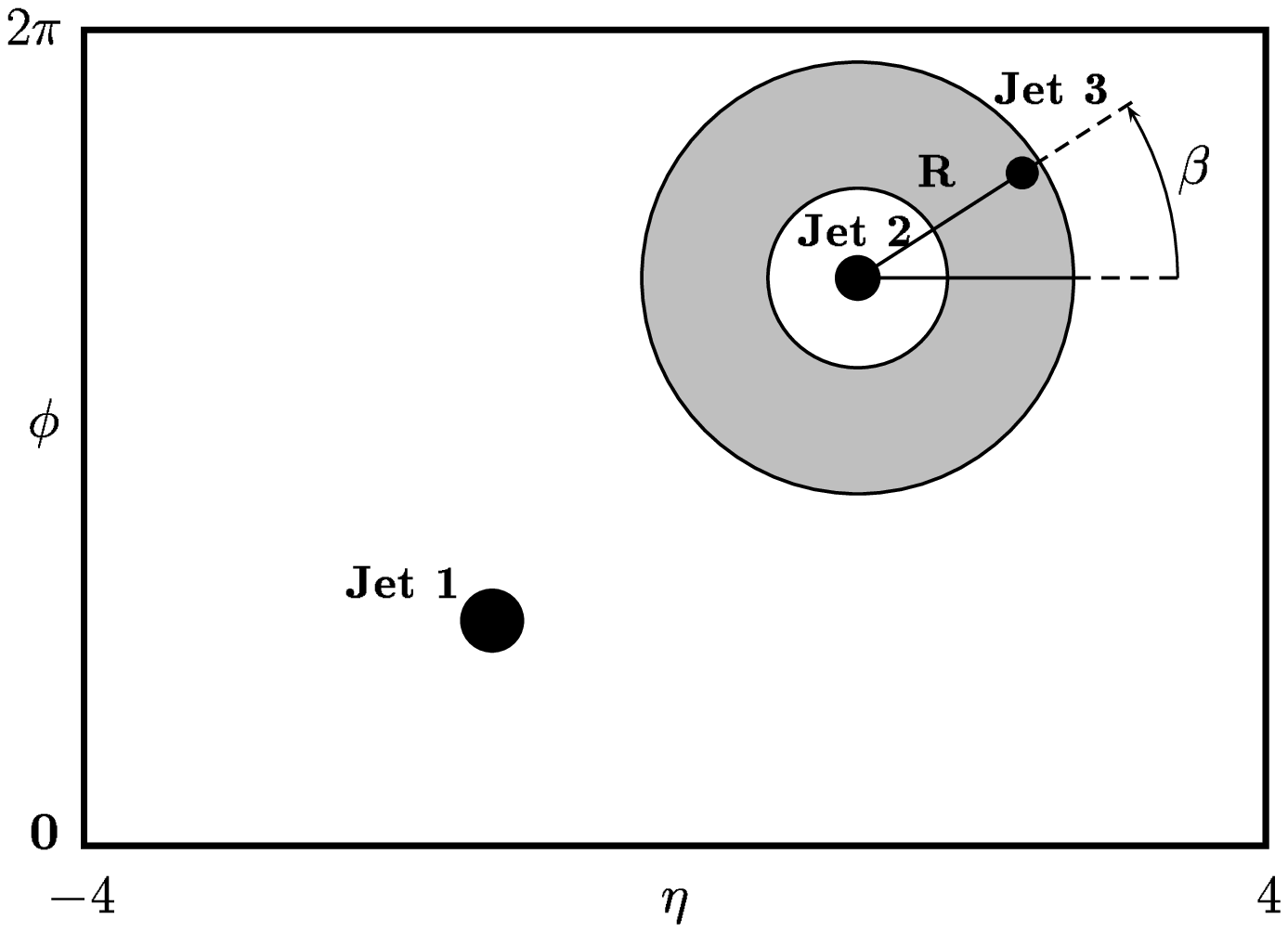,height=5.9cm,width=6.5cm}}
 \vspace{2.cm}
 \label{PL_FIG2}
 \vspace{0.1cm}
\end{figure}

\newpage
\vskip 0.5in
{\centerline {\bf Figure 3}}
\vskip 1in

\begin{figure} 
 \centerline{\psfig{figure=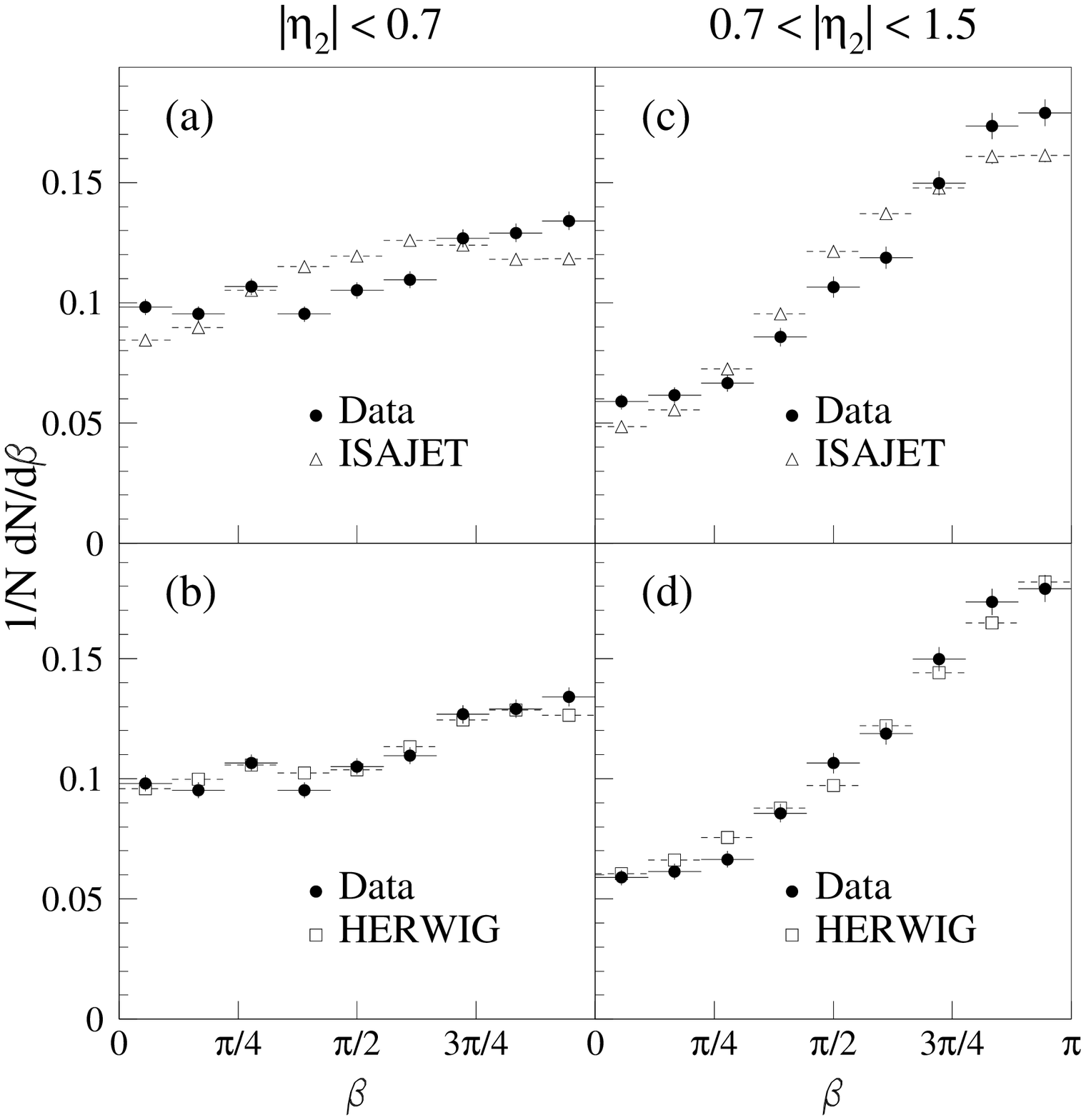,width=14cm}}
 \vspace{0.6cm}
 \label{PL_FIG3}
 \vspace{0.1cm}
\end{figure}

\newpage
\vskip 0.5in
{\centerline {\bf Figure 4}}
\vskip 1in

\begin{figure} 
 \vspace{-0.2cm}
 \centerline{\psfig{figure=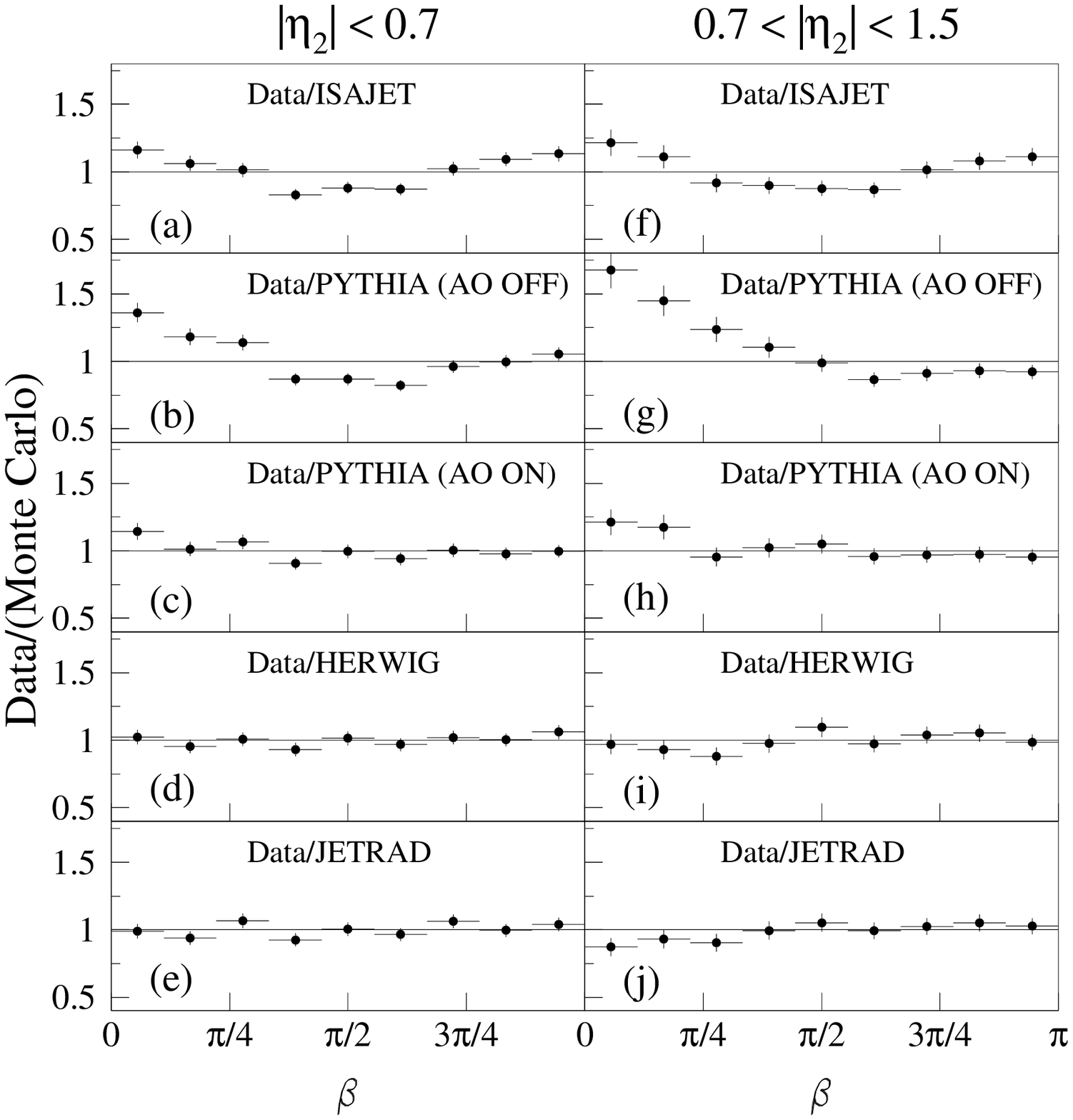,width=15cm}}
 \vspace{0.9cm}
 \label{PL_FIG4}
 \vspace{0.1cm}
\end{figure}


\begin{references}

%
\bibitem[*]{ecuador}
Visitor from Universidad San Francisco de Quito, Quito, Ecuador.

\bibitem[\dag]{beijing}
Visitor from IHEP, Beijing, China.

\vskip 0.25cm

\bibitem{cc1} JADE Collaboration, W. Bartel {\sl et al.}, Phys. Lett. 
              {\bf B101}, 129
              (1981); Zeit. Phys. {\bf C21}, 37 (1983); Phys. Lett. {\bf B134},
              275 (1984); Phys. Lett. {\bf B157}, 340 (1985).
\bibitem{cc2} TPC/2$\gamma$ Collaboration, H. Aihara {\sl et al.}, Phys. Rev. 
              Lett.
              {\bf 54}, 270 (1985); Zeit. Phys. {\bf C28}, 31 (1985); Phys. Rev.
              Lett. {\bf 57}, 945 (1986).
\bibitem{cc3} TASSO Collaboration, M. Althoff {\sl et al.}, Zeit. Phys. 
              {\bf C29}, 29 (1985).
\bibitem{cc4} MARK2 Collaboration, P.D. Sheldon {\sl et al.}, Phys. Rev. Lett. 
              {\bf 57}, 1398 (1986).
\bibitem{cc5} OPAL Collaboration, M.Z. Akrawy {\sl et al.}, Phys. Lett. 
              {\bf B247},
              617 (1990); Phys. Lett. {\bf B261}, 334 (1991); P.D. Acton
              {\sl et al.}, Phys. Lett. {\bf B287}, 401 (1992); Zeit. Phys. 
              {\bf C58}, 207 (1993).
\bibitem{cc5a} L3 Collaboration, M. Acciarri {\sl et al.}, Phys. Lett. 
              {\bf B353}, 145 (1995).
\bibitem{cc6} B. Andersson, G. Gustafson, and T. Sj\"{o}strand, Phys. Lett. 
              {\bf B94}, 211 (1980).
\bibitem{cc7} Ya.I. Azimov, Yu.L. Dokshitzer, V.A. Khoze, and S.I. Troyan,
              Phys. Lett. {\bf B165}, 147 (1985); Sov. Journ. Nucl. Phys. {\bf
              43}, 95 (1986).
\bibitem{cc8} R.K. Ellis, G. Marchesini, and B.R. Webber, Nucl. Phys.  
              {\bf B286}, 643 (1987); Erratum Nucl. Phys. {\bf B294}, 1180 
              (1987).
\bibitem{cc9} Yu.L. Dokshitzer, V.A. Khoze, A.H. Mueller, and S.I. Troyan, 
              Basics of Perturbative QCD, Editions Fronti\`{e}res (1991); Rev.
              Mod. Phys. {\bf 60}, 373 (1988); Yu.L. Dokshitzer, V.A. Khoze, and
              S.I. Troyan, Sov. Journ. Nucl. Phys. {\bf 46}, 712 (1987).  
\bibitem{cc10} CDF Collaboration, F. Abe {\sl et al.}, Phys. Rev. D {\bf 50},
               5562 (1994).
\bibitem{cc11} \D0\ Collaboration, S. Abachi {\sl et al.}, Nucl. Instrum. Meth. 
               {\bf A338}, 185 (1994).
\bibitem{cc12}  D. Cullen-Vidal, Ph.D. Dissertation, Brown University, 
               (1997) (unpublished).
\bibitem{esca} \D0\ Collaboration, R. Kehoe, in Proceedings of the VIth 
               International 
               Conference on Calorimetry in High Energy Physics,
               Laboratori Nazionali di Frascati, Frascati, Italy, (1996).
\bibitem{cc13} F.E. Paige and S.D. Protopopescu, BNL report No. 38034, 1986
               (unpublished).
\bibitem{cc14} G. Marchesini {\sl et al.}, Computer Physics Commun. {\bf 67}, 
               465 (1992).
\bibitem{cc16} T. Sj\"{o}strand, Computer Physics Commun. {\bf 82}, 74 (1994).
\bibitem{cc15} W.T. Giele, E.W.N. Glover, and D.A. Kosower, Nucl. Phys. 
               {\bf B403}, 633 (1993); Phys. Rev. Lett. {\bf 73}, 2019 (1994).
\bibitem{cteq} CTEQ Collaboration, H.L. Lai {\sl et al.}, Phys. Rev. D 
               {\bf 51}, 4763 (1995).
\bibitem{geant} R. Brun {\sl et al.}, ``GEANT 3.14'' (unpublished), CERN, 
                DD/EE/84-1.

\end{references}
\end{document}